\documentclass[onecolumn,prl,showkeys]{revtex4}
\usepackage{amsmath}
\usepackage{amsfonts}
\usepackage{epsfig}
\usepackage[colorlinks,hyperindex]{hyperref}

\newcommand{\eq}[1]{Eq.~\ref{#1}}
\newcommand{\Figref}[1]{Fig.~\ref{#1}}
\newcommand{\Tabref}[1]{Tab.~\ref{#1}}

\newcommand{\gev}{\:\text{GeV}}
\newcommand{\tev}{\:\text{TeV}}

\def\bea{\begin{eqnarray} }
\def\eea{ \end{eqnarray} } 
\def\be{\begin{equation} }
\def\ee{ \end{equation} }

\begin{document}

\title{Where to look for natural supersymmetry}

\author{S.~S.~AbdusSalam}
\email{Shehu.AbdusSalam@Roma1.infn.it}
\affiliation{INFN, Sezione di Roma, Piazzale Aldo Moro, 2, I-00185
  Roma, c/o Dipartimento di Fisica - Universit\'a degli Studi di Roma
  ``La Sapienza'', Italy}  
\author{L.~Velasco-Sevilla}
\email{Liliana.Velasco-Sevilla@uib.no}
\affiliation{ University of Bergen, Department of Physics and
  Technology,\\ PO Box 7803, 5020 Bergen, Norway} 
\begin{abstract}
Why is natural supersymmetry neither detected nor ruled-out to date? To
answer this question we use the Bayesian approach where the emphasis
in finding prior-independent features within broader and minimally
biased frames is taken as the guiding principle. The 20-parameter
minimal supersymmetric standard model (MSSM) global fits to subjective
naturalness indicate the existence of a prior-independent upper bound
on the pseudoscalar Higgs boson mass $m_A$ as a function of
$\tan\beta$, the ratio of the vacuum expectation values of MSSM Higgs
doublets. For a 30-parameter MSSM this implies that
$m_A \lesssim 3\tev$ and $\tan \beta \lesssim 25$ at 95\% Bayesian
confidence. Removing the contradictory subjectiveness within the
electroweak fine-tuning measure leads to finding the naturalness line,
$m_A \sim \frac{1}{\sqrt{2}} \, m_Z \, \tan \beta,$ that reduces by
one the number of MSSM Higgs sector free parameters. 
\end{abstract}


\maketitle

\paragraph{Introduction:} Supersymmetry \cite{Martin:1997ns} model
constructions and phenomenological studies go decades back 
\cite{Wess:1974tw, Fayet:1976et,Fayet:1977yc,
  Dimopoulos:1981zb, Nilles:1983ge, Haber:1984rc, Barbieri:1982eh,
  Dawson:1983fw} but yet have not been discovered  
nor ruled out by high-energy physics experiments. The
specific prediction from supersymmetry is that there must be new,
beyond the standard model, particles with and without colour
charges. But it does not specify what the particular masses and
couplings of the new particles will be. These remain arbitrary with
more than 100 free parameters. Currently, on the experiments side, it is   
expected that the large hadron collider (LHC) will give a definite
answer as to whether low-energy supersymmetry has any role in
stabilising the Higgs boson mass at $125 \gev$ \cite{Aad:2012tfa,
  Chatrchyan:2012ufa, Aad:2015zhl}. It is going to probe the so-called ``natural''
supersymmetry. {{In view of this we ask the question:}}
 Is there any robust prediction from  
natural supersymmetry that could be targeted by the experiments for
a discovery or an absolute exclusion? We are going to address this 
question by employing Bayesian statistical 
techniques. The Bayesian method can be used to extract robust
predictions from a model based on experimental data. Here the model
under consideration will be the R-parity conserving minimal supersymmetric
standard model (MSSM).

Robust predictions can be extracted because the Bayesian approach
allows a check for the stability of results with respect to widely
different but well-motivated changes in the prior assumptions
concerning the base supersymmetry parameters. The results which remain 
the same under the change of the prior assumptions are said to be
prior-independent and represent the predictions by the model based on
the experimental data. Before the LHC
commissioning, the 20-parameter MSSM fits 
\cite{AbdusSalam:2008uv,AbdusSalam:2009qd} to neutralino cold dark
matter (CDM) relic 
density, electroweak and B-physics data revealed two observables to be
prior-independent, namely the (then undiscovered) Higgs boson and the
lightest top-squark masses. The Higgs boson mass was predicted to
lie between 119 and 128 GeV within the 95\% Bayesian credibility
region, while the top-squark mass to be around $2 \tev$. The
20-parameter MSSM is specified by 
\be \label{20par}
\underline{\theta} = \{ M_{1,2,3};\;\; m^{3rd \, gen}_{\tilde{f}_{Q,U,D,L,E}},\;\; 
m^{1st/2nd \, gen}_{\tilde{f}_{Q,U,D,L,E}}; \;\;A_{t,b,\tau,\mu=e},
\;\;m^2_{H_{u,d}}, \;\;\tan \beta; \;\; m_Z, \;\;m_t, \;\;m_b,
\;\;\alpha_{em}^{-1}, \;\;\alpha_s \}
\ee
where the gaugino mass parameters $M_1$, $M_2$ and $M_3$
were allowed in -4 to 4 TeV range.  The sfermion $\tilde f$ mass
parameters $m_{\tilde 
  f}$ vary between 100 GeV to 4 TeV. The trilinear scalar couplings
$A_{t,b,\tau,\mu=e} \in [-8, 8]$ TeV. The Higgs-sector parameters 
$m^2_{H_u}$, $m^2_{H_d}$, were varied according to $m^2 \in
sign(m)\,[-4, 4]^2 \textrm{TeV}^2$. The ratio of the vacuum
expectation values $\tan \beta=\left<H_u\right>/\left<H_d\right>$  is
allowed to be between 2 
and 60, while $sign(\mu)$ the sign of the Higgs doublets mixing
parameter, is allowed to be randomly $\pm 1$.  The remaining five 
standard model parameters were also varied in a Gaussian manner with
central values and deviations according to experimental results
\cite{pdg}.

In this article we are going to show that by imposing fine-tuning cuts
within the 20-parameters MSSM, an additional prior-independent result
manifests. From this, an inequality relation between the pseudoscalar
Higgs boson mass $m_A$ and $\tan\beta$ can be deduced. For the
cuts we use the electroweak fine-tuning measure \cite{Baer:2012up,
  Baer:2012cf} $\Delta_{EW}$ defined as follows. Consider the electroweak symmetry
breaking condition for a 1-loop corrected Higgs potential,
$V + \Delta V$  
\be \label{Hmin1} \frac{m_Z^2}{2} = \frac{m_{H_d}^2 + \Sigma_d^d - 
    (m_{H_u}^2+\Sigma_u^u)\tan^2\beta}{\tan^2\beta -1} -\mu^2.\ee
Here $\Sigma_u^u$ and $\Sigma_d^d$ arise from the 1-loop radiative
corrections. For naturalness, each term in the right hand side of 
\eq{Hmin1} should be comparable to $m^2_Z/2$ so that 
\be \label{DEW0} \Delta_{EW} \equiv max_i \left(C_i\right) /
(m_Z^2/2) \ee
accommodates the fact that for obtaining a natural value of
$m_Z$ then the terms $C_i$, with $i=H_d,\ H_u$, $\mu$,
$\Sigma_u^u(k)$, $\Sigma_d^d(k)$, where $k$ denotes the various
particles and sparticles contributions, must be of order $m_Z^2/2$.
Using the terms that couple the most to the Higgs sector (the case
$k = \tilde{t}_{1,2}, \tilde{b}_{1,2}$) we have 
\be \begin{split} 
  C_\mu &= |-\mu^2|, \\
  C_{H_u} &= |-m_{H_u}^2\tan^2\beta /(\tan^2\beta -1)|, \,\\
  C_{H_d} &= |m_{H_d}^2/(\tan^2\beta -1)|,\\
  C_{\Sigma^d_d} &= | \Sigma^d_d/(\tan^2\beta -1)|, \\
  C_{\Sigma^u_u} &= | -\Sigma^u_u \, \tan^2 \beta/(\tan^2\beta -1)|, \\
  \Sigma^{d,u}_{d,u} &= \Sigma_{i} \, | \Sigma^{d,u}_{d,u} (i)|.
\end{split} 
\ee 
The expressions for $\Sigma^{d,u}_{d,u} (i)$ are shown in the
Appendix. 

In the next section, we describe the Bayesian approach to MSSM
naturalness, the fitting procedure and the prior-independent result
obtained. After that we explain the impact of 
the result which is a prior-independent bound on $m_A$ as a function
of $\tan \beta$ on the a 
30-parameters MSSM posterior distribution. We then assess to
what extent has some relevant 8 TeV LHC supersymmetry limits probe the
natural MSSM-30. At the end, we present an analytical argument that 
exposes a subtle methodological contradiction by looking closer at the
electroweak fine-tuning measure. Fixing the contradiction lead to
a no fine-tuning ``naturalness line''. After this we summarise our
results and give an outlook for future studies. 

\paragraph{Naturalness, the Bayesian approaches:} There are two major
trends in the literature concerning Bayesian approach to MSSM
naturalness. First, for addressing
MSSM naturalness one can compute the amount of 
fine-tuning at each point during the parameters sampling and then
penalise highly fine-tuned points according to a chosen subjective
limit (see e.g. \cite{Allanach:2006jc}). Within this method, various
groups use different fine-tuning measures,
e.g. \cite{Harnik:2003rs,Kitano:2005wc,Ellis:1986yg,Barbieri:1987fn}. The
difference measures, however, agree when used appropriately as
explained in \cite{Baer:2014ica,Baer:2013gva}. According to the second
trend, fine-tuning measures manifest implicitly within the Bayesian
global fit procedures. In
\cite{Cabrera:2008tj,Ghilencea:2012qk,Fowlie:2014xha} it is shown that 
fitting the MSSM parameters in a Bayesian way automatically
incorporate a fine-tuning penalisation. Our approach in this article
goes along the first trend. We use the electroweak fine-tuning measure
\eq{DEW0} and penalise or rule-out MSSM points with $\Delta_{EW}
>4$. The choice $\Delta_{EW} >4$ in search for prior-independent
results from global fits to MSSM represents the ``naturalness''
data. A natural MSSM point should have $\Delta_{EW} \rightarrow 1.$
Relaxing away from $\Delta_{EW} = 1$ as a fine-tuning cut we choose
$\Delta_{EW} < 2 + 2$ where the first ``2'' represents a 50\%
fine-tuning and the second a 100\% ``theoretical'' allowance on the
first. The Bayesian global fit procedure with $\Delta_{EW} \leq 4$
is described as follows. 

\paragraph{Fitting procedure:} Based on the methodology for our MSSM
programme \cite{Feroz:2008wr, AbdusSalam:2008uv, 
  AbdusSalam:2009qd, AbdusSalam:2009tr, AbdusSalam:2010qp,
  AbdusSalam:2011hd, AbdusSalam:2012sy, AbdusSalam:2012ir,
  AbdusSalam:2013qba} the Bayesian global fit of the 20-parameters
MSSM plus 5 standard model parameters (MSSM-25) were
performed separately with linear and logarithmic prior probability
distributions on the parameters \eq{20par}. These were
fit to the Higgs boson mass, naturalness requirement, neutralino CDM
relic  density, electroweak and B-physics data shown 
in 
\Tabref{tab:obs}. 
\begin{table} 
\begin{center}{\begin{tabular}{|cl||cl|}
\hline
Observable & Constraint & Observable & Constraint  \\ 
\hline
$m_W$ [GeV]& $80.399 \pm  0.027$ \cite{verzo}&$A^l = A^e$& $0.1513 \pm
0.0021$ \cite{:2005ema} \\
$\Gamma_Z$ [GeV]& $2.4952 \pm 0.0025$ \cite{:2005ema}&$A^b$ & $0.923
\pm 0.020$ \cite{:2005ema}\\ 
$\sin^2\, \theta_{eff}^{lep}$  & $0.2324 \pm 0.0012$ \cite{:2005ema}&$A^c$ & $0.670 \pm 0.027$ \cite{:2005ema}\\  
$\delta a_\mu $ & $(30.2 \pm 9.0) \times 10^{10}$
\cite{Bennett:2006fi,Davier:2007ua} &$Br(B\rightarrow
X_s \gamma)$ & $(3.55 \pm 0.42) \times 10^{4}$ \cite{Barberio:2007cr}\\  
$R_l^0$ & $20.767 \pm 0.025$ \cite{:2005ema} &$Br(B_s \rightarrow \mu^+ \mu^-)$ & $
3.2^{+1.5}_{-1.2} \times 10^{-9}$ \cite{Aaij:2012nna}\\  
$R_b^0$ & $0.21629 \pm 0.00066$ \cite{:2005ema}&$R_{\Delta M_{B_s}}$ & $0.85 \pm 0.11$\cite{Abulencia:2006ze}\\ 
$R_c^0$ & $0.1721 \pm 0.0030$ \cite{:2005ema}&$R_{Br(B_u \rightarrow \tau \nu)}$&
$1.26 \pm 0.41$ \cite{Aubert:2004kz,paoti,hep-lat/0507015}\\ 
$A_{\textrm{FB}}^b$ & $0.0992 \pm 0.0016$ \cite{:2005ema}&$\Delta_{0-}$ & $0.0375 \pm
0.0289$\cite{J.Phys.G33.1}\\  
$A_{\textrm{FB}}^c$ & $0.0707 \pm 0.035$ \cite{:2005ema}&$\Omega_{CDM} h^2$ & $0.11
\pm 0.02 $ \cite{0803.0547}\\ 
 & & $m_h$ & $125.6 \pm 3.0$  [GeV]\cite{ATLAS:2013mma, CMS:yva}\\ 
\hline
\end{tabular}}\end{center}
\caption{Summary for the central values and errors for the Higgs boson
  mass, the electroweak physics observables, B-physics observables and
  cold dark matter relic density constraints.\label{tab:obs}}  
\end{table}
%
{\sc MultiNest}~\cite{Feroz:2007kg, Feroz:2008xx}  
package which implements nested sampling algorithm~\cite{Skilling} for
exploring model parameters space were used. At each MSSM-25 point the
supersymmetry spectra were computed via {\sc
  SOFTSUSY}~\cite{Allanach:2001kg} and the list of observables $O_i$,
\be
\begin{split} \label{alldata}
\underline O = &\{m_W,\; \sin^2\, \theta^{lep}_{eff},\; \Gamma_Z,\;
\delta 
a_{\mu},\; R_l^0,\; A_{fb}^{0,l},\; A^l = A^e,\; R_{b,c}^0,\;
A_{fb}^{b,c},\; A^{b,c},\;  BR(B \rightarrow X_s \, \gamma),\; \\
& BR(B_s\rightarrow \mu^+ \, \mu^-), \; \Delta_{0-},\; R_{BR(B_u
  \rightarrow \tau \nu)},\; R_{\Delta M_{B_s}}, \Omega_{CDM}h^2,\;
m_h, \, \Delta_{EW}^{-1} \geq 5\% \, \}, 
\end{split}
\ee 
via the following packages. {\sc
  micrOMEGAs}~\cite{Belanger:2008sj} was used for computing
neutralino CDM relic density $\Omega_{CDM}h^2$ and the anomalous
magnetic moment of the muon $\delta
a_\mu$; and {\sc SuperIso}~\cite{Mahmoudi:2007vz} for predicting 
$BR(B_s \rightarrow \mu^+ \mu^-)$, $BR(B \rightarrow s \gamma)$ and
the isospin asymmetry, $\Delta_{0-}$, in $B \rightarrow K^* \gamma$. 
With {\sc susyPOPE}~\cite{Heinemeyer:2006px,Heinemeyer:2007bw} we 
computed the $W$-boson mass $m_W$, the effective leptonic mixing
angle variable $\sin^2  \theta^{lep}_{eff}$, the total $Z$-boson decay
width, $\Gamma_Z$, and the other electroweak observables. These allow
the computation of the posterior probability via Bayes' theorem,
\be \label{posterior}
 p(\underline \theta |\underline d, {\cal{H}}) = L_{\Delta_{EW}} \, L_{CDM}(x) \prod_i \, \frac{
   e^{\left[- (O_i - \mu_i)^2/2 \sigma_i^2\right]}}{\sqrt{2\pi
     \sigma_i^2}} \, \frac{p(\underline \theta | {\cal{H}})}
 {p(\underline d | {\cal{H}})}; 
\ee
\be
L_{\Delta_{EW}} = 
\begin{cases}
1, & \textrm{if $\Delta_{EW}^{-1} \geq 50\%$} \\  
0, & \textrm{if $\Delta_{EW}^{-1} < 50\%$} \\
\end{cases},
\quad
L_{CDM}(x) = 
\begin{cases}
1/(y + \sqrt{\pi s^2/2}), & \textrm{if $x < y$} \\  
e^{\left[-(x-y)^2/2s^2\right]}/(y + \sqrt{\pi s^2/2}),
& \textrm{if $x \geq y$} \\
\end{cases}. 
\ee
Here $i$ run over the different experimental observables
(data) other than the CDM relic density, $x$ represents the predicted
value of the neutralino CDM relic density, $y = 0.11$ is the WMAP central
value quoted in \Tabref{tab:obs} and $s=0.02$ the inflated
error. The likelihood contribution coming from the CDM relic density
is given by $L_{CDM}(x)$ which is purely Gaussian when the predicted relic
density $x$ is greater than the experimental central value $y =
0.11$ thus imposing penalisation for CDM over-production. No
penalisation is imposed when $x < y$. The set of experimental data
used for the fits is
\be
\underline{d} = d_{\Tabref{tab:obs}} + d_{\Delta_{EW}} = \{ \mu_i,
  \sigma_i \} +  \{ \Delta_{EW}^{-1} > 50 \% \}.
\ee
Here $d_{\Tabref{tab:obs}}$ is the set of experimental central values
$\mu_i$ and error $\sigma_i$ shown in \Tabref{tab:obs}. ${\cal H}$ in
\eq{posterior} represents the context or hypothesis for the Bayesian
theorem. i.e. nature is supersymmetric and that neutralinos make part
of the cold dark matter relics. From the posterior of the
global fits we only show the result which is approximately
prior-independent. This happens to be an MSSM-25  feature in the
$(m_A, \, \tan \beta)$ plane. 

\begin{figure*}[!b]
  \includegraphics[width=5.5cm]{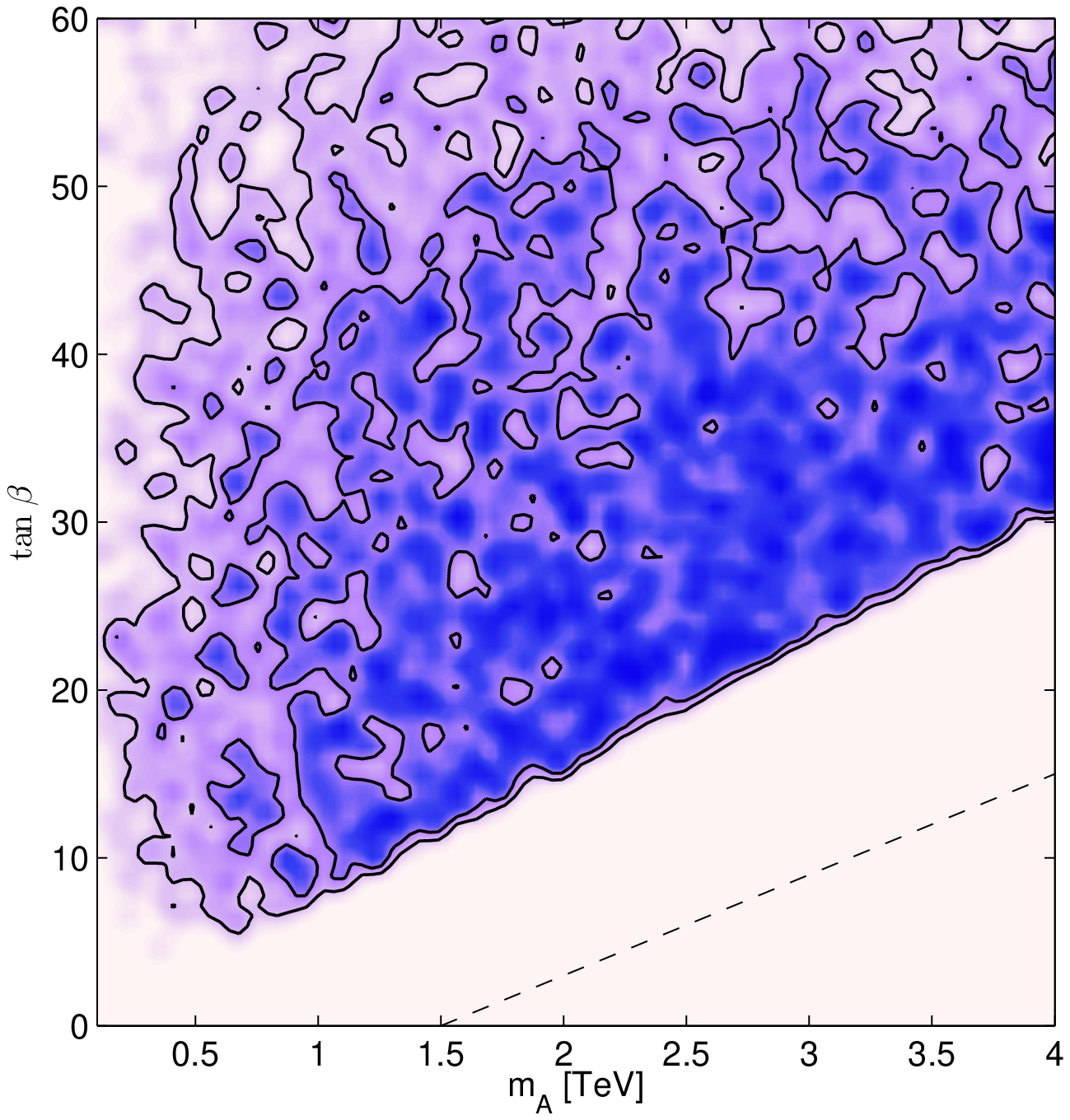}
  \includegraphics[width=5.5cm]{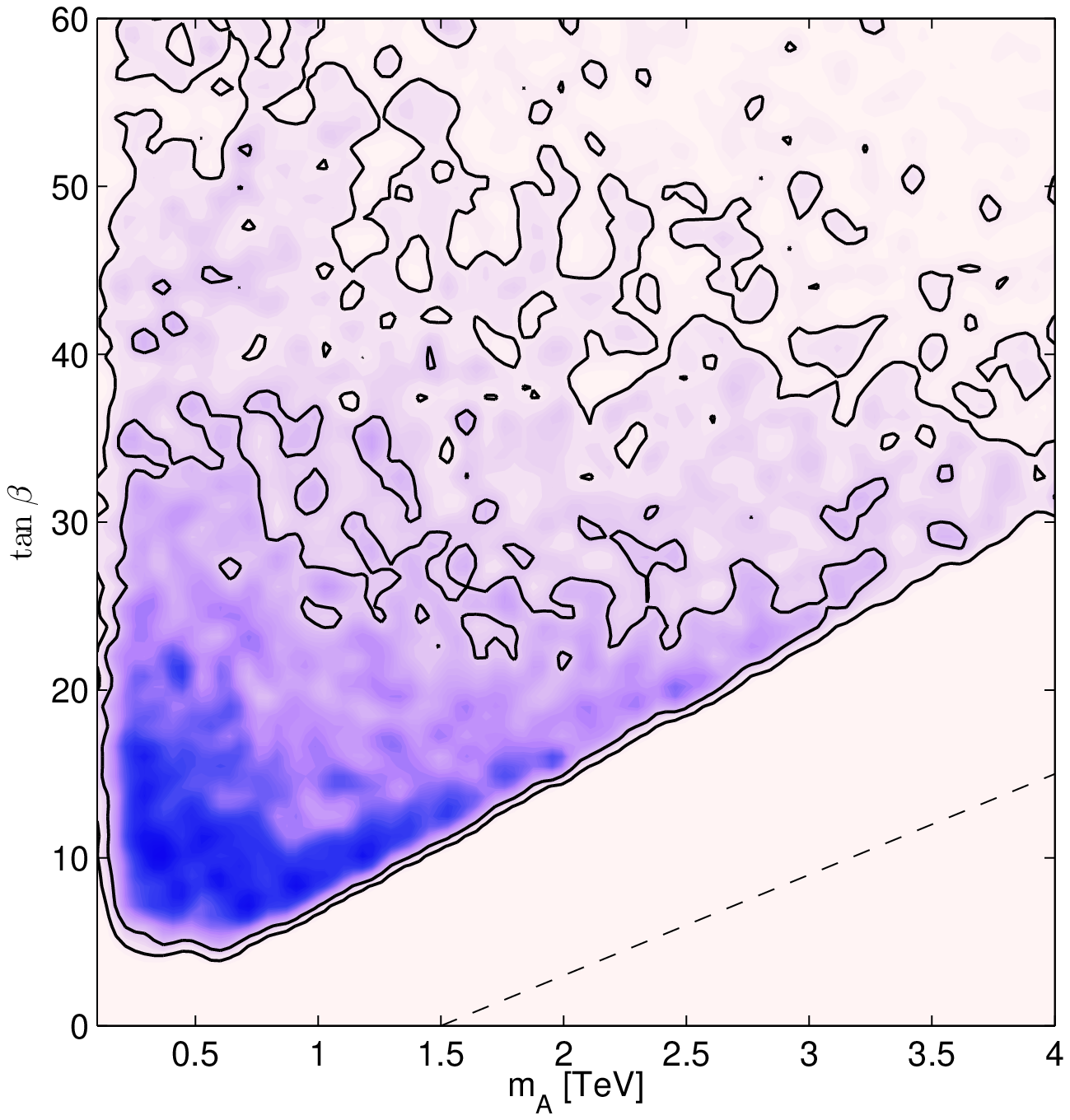}
  \caption{Two-dimensional posterior distributions from the
    MSSM-25 fits to experimental plus ``naturalness'' data. The
    left-side (right-side) 
    plot is for logarithmic (flat) prior fit. The empty triangular
    region on the plots are explicitly excluded by the naturalness
    limit $\Delta_{EW} \leq 4$. The dashed line represents the shift
    that will occur when a relaxed naturalness cut $\Delta_{EW} \leq
    20$ is imposed. The solid contour lines enclose the dark blue
    (dark) and light violet (light grey to white) regions which
    correspond respectively to the 68\% and 95\% Bayesian
    credibility. For both panels dark blue (dark) regions have
    higher probability compared to the light blue to light violate
    (grey to white) ones.}
  \label{fig:mAtb_0}
\end{figure*}

\paragraph{Result:} The two-dimensional
posterior distributions in \Figref{fig:mAtb_0}  
shows that requiring fine-tuning no worse than 50\% as 
naturalness data while fitting the MSSM-25 to data has a
prior-independent impact
in the $(m_A, \tan \beta)$ plane. The empty triangular regions are
excluded by this naturalness requirement. The prior-independent result
is 
\be \label{cutfromfit} m_A <  \, \frac{4}{30} \, \tan \beta \,
\tev. \ee
\eq{cutfromfit} is robust and can be applied to any
supersymmetry model with a necessary electroweak symmetry breaking
condition \eq{Hmin1}. Next, we assess the impact of this on the posterior
sample of an MSSM-30 which favours low values of $\tan \beta \lesssim
27$ within 95\% Bayesian credibility. First we give a brief introduction
of the MSSM-30 frame and then afterwards check the natural
(\eq{cutfromfit}-based) MSSM-30 points against some LHC
supersymmetry limits. 

\paragraph{Naturalness constraint on MSSM-30:} In
\cite{AbdusSalam:2014uea}, the 30-parameters MSSM was constructed by
reducing the parent 100+ MSSM parameters using a systematic treatment
of minimal flavour violation -- unlike as done by hand for the MSSM-25
case. The 
parameters consist of $e^{\phi_1} M_1$, $e^{\phi_2} M_2$,
and $M_3$ in the gaugino sector with $M_{1,\, 2}$ (and also their
imaginary parts $Im(M_{1,\, 2})$) which are varied between -4 to
4 TeV. $M_3$ is allowed to be between 100 GeV to 4 TeV. Within the
Higgs sector, $m_A$ is varied between $100 \gev$ to $4 \tev$ while
$\mu$ and $Im(\mu)$ were allowed
within -4 to 4 TeV. As for MSSM-25, $\tan \beta$  is allowed to be
between 2 and 60. The scalar mass and trilinear coupling parameters
are 
\noindent
\begin{minipage}{.48\linewidth}
  \bea \label{mfvpar30}
  & &M^2_Q = \tilde{a}_1 + x_1 X_{13} + y_1 X_1,\quad M^2_E = \tilde{a}_7 + y_7 X_1, \nonumber\\ \nonumber
  & &M^2_U = \tilde{a}_2  + x_2 X_1,           \quad A_E = \tilde{a}_8 X_1,\\\nonumber 
  & &M^2_D = \tilde{a}_3 + y_3 X_1,            \quad A_U = \tilde{a}_4 X_5 + y_4 X_1,\\\nonumber
  & &M^2_L = \tilde{a}_6 + y_6 X_1,            \quad A_D = \tilde{a}_5 X_1 + y_5 X_5,\\\nonumber
  \eea
\end{minipage}%
\begin{minipage}{.48\linewidth}
  \begin{eqnarray}
    \label{xbasis}
    \begin{array}{cccc}
      X_1 = \delta_{3i} \delta_{3j}, & X_2 = \delta_{2i} \delta_{2j}, &
      X_3 = \delta_{3i} \delta_{2j}, & X_4 = \delta_{2i} \delta_{3j},
      \\
      X_5 = \delta_{3i} V_{3j}, & X_6 = \delta_{2i} V_{2j}, & X_7 =
      \delta_{3i} V_{2j}, &X_8 = \delta_{2i} V_{3j}, \\
      X_9 = V^*_{3i} \delta_{3j}, & X_{10} = V^*_{2i} \delta_{2j}, &
      X_{11} = V^*_{3i} \delta_{2j}, & X_{12} = V^*_{2i} \delta_{3j},
      \\
      X_{13} = V^*_{3i} V_{3j}, & X_{14} = V^*_{2i} V_{2j}, & X_{15} =
      V^*_{3i} V_{2j}, & X_{16} = V^*_{2i} V_{3j}.\\
    \end{array}
  \end{eqnarray}
\end{minipage}

\begin{figure*}[!t]
  (a)~\includegraphics[width=5.5cm]{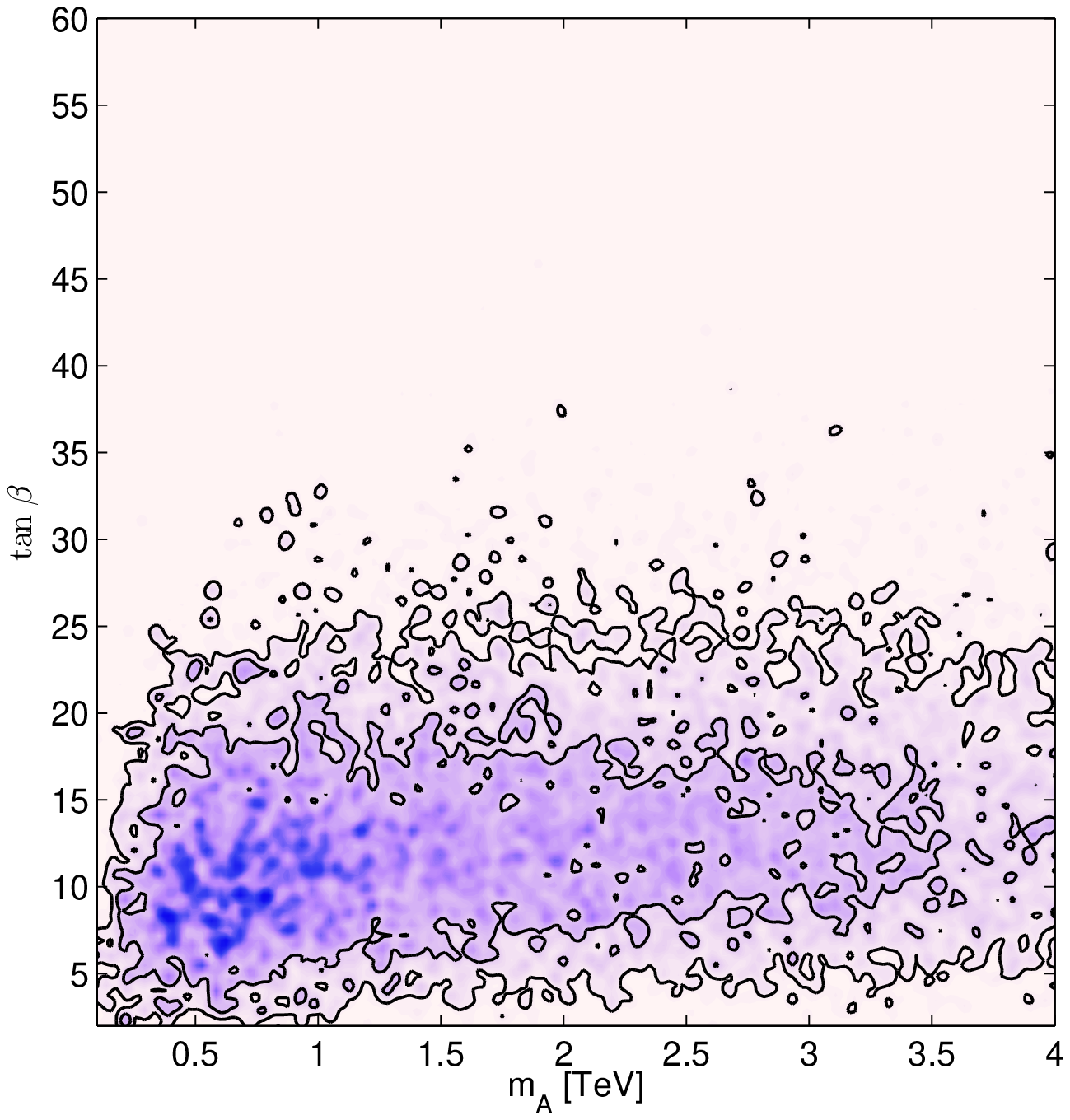}     
  (b)~\includegraphics[width=5.5cm]{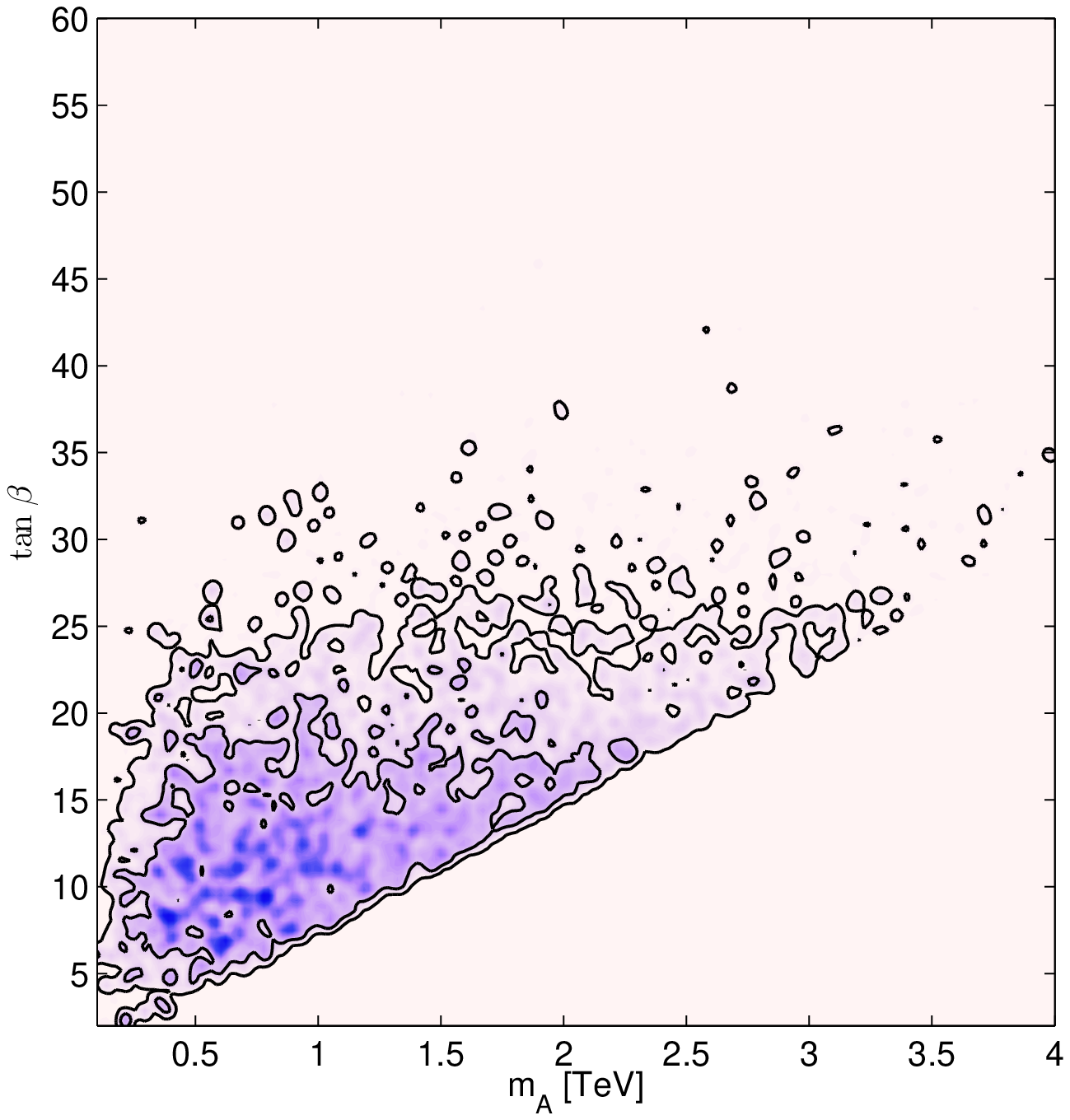} 
  (c)~\includegraphics[width=5.5cm]{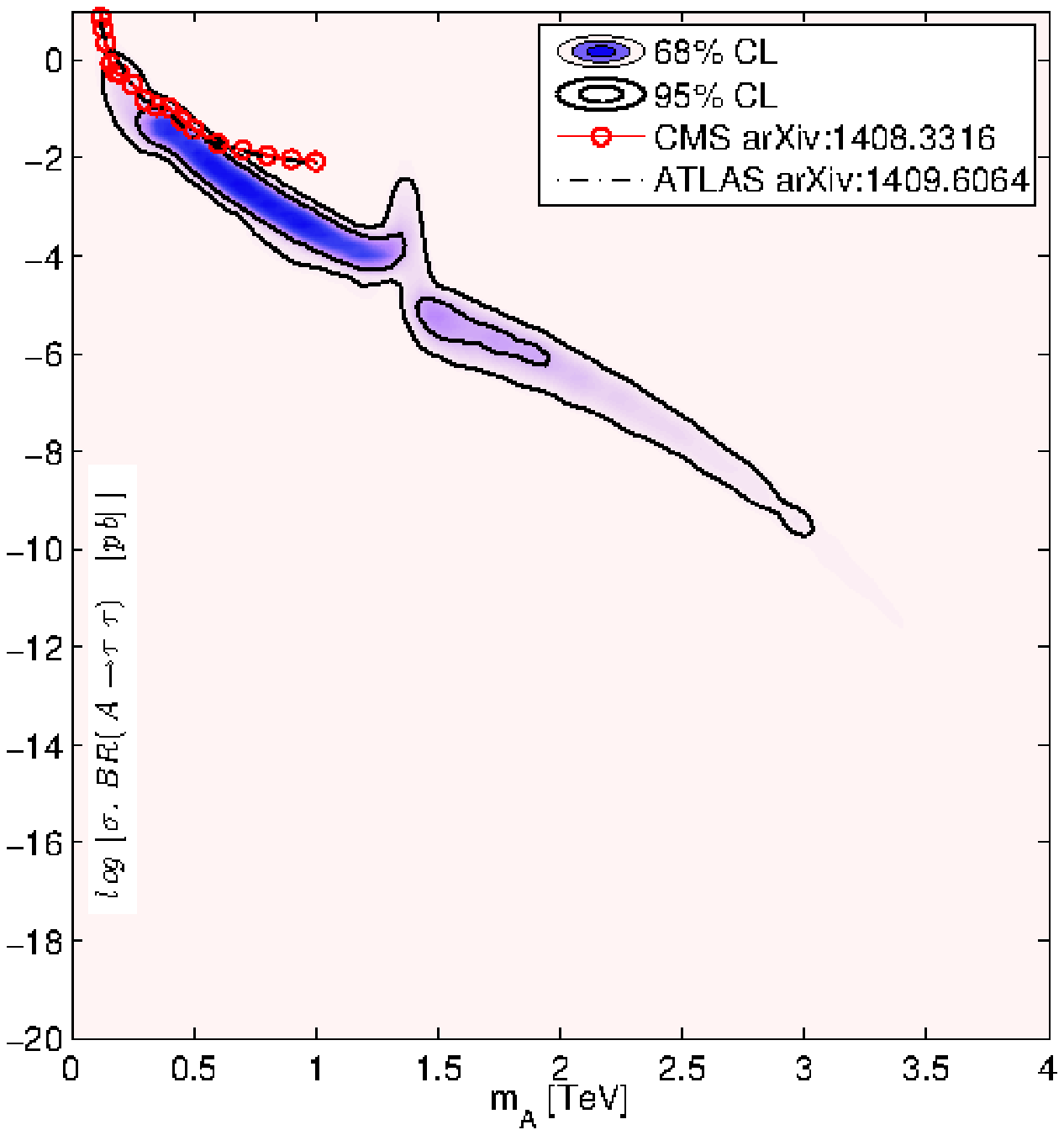}     
  (d)~\includegraphics[width=5.5cm]{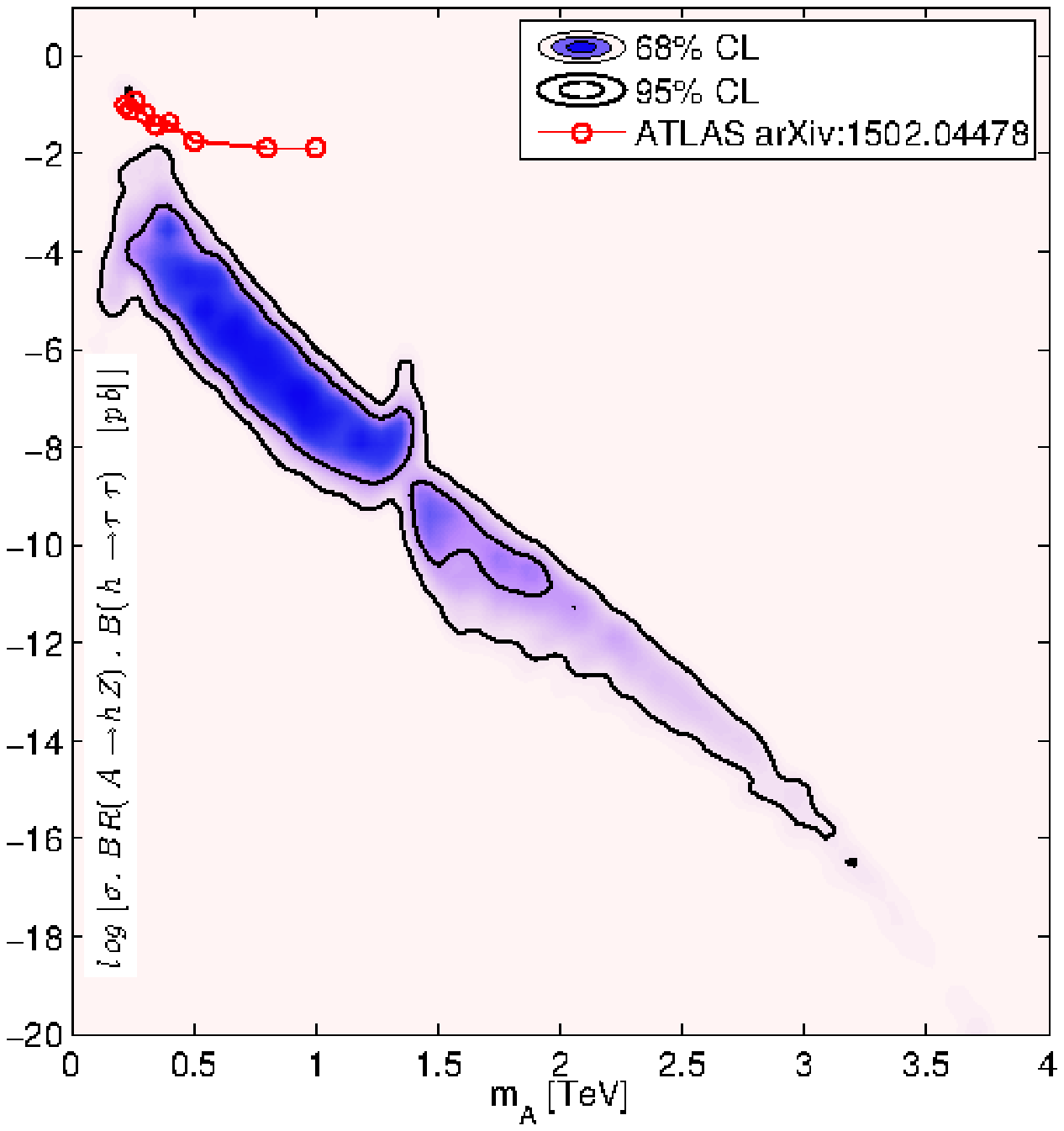} 
  \caption{(a) The posterior distribution for the MSSM-30
    on ($m_A$, $\tan \beta$) plane. (b)
    Explicitly shows the effect of the naturalness cut
    \eq{cutfromfit} on the MSSM-30 posterior sample. Note that small patches of regions that
    show up in (b) but absent in the parent plot (a) is an artifact of colouring and contouring 
    interpolations/normalisation. (c) and (d)
    show the effect of the ATLAS  
    and CMS 95\% upper bounds on the production cross-section times
    decay branching ratios for the MSSM-30 pseudoscalar Higgs
    boson. For all the panels dark blue (dark) regions have
    higher probability compared to the light blue to light violate
    (grey to white) ones.} 
  \label{fig:mAtb}
\end{figure*}

\noindent
The bases $X_{1, \ldots, 16}$ are products amongst Kronecker delta
$\delta$ and the Cabibbo-Kobayashi-Maskawa mixing matrix $V$
elements. The parameters $\tilde{a}_{1,2,3,6,7} > 0$ and $x_{1,2},
y_{1,3,6,7}$ were varied within $(100 \gev)^2$ to $(4 \tev)^2$ and $-(4
\tev)^2$ to $(4 \tev)^2$ respectively; while $\tilde{a}_{4,5,8}$,
$Im(\tilde{a}_{4,5,8})$, $y_{4,5}$ and $Im(y_{4,5})$ were allowed
between $-8 \tev$ to $8 \tev$. The SM parameters are fixed according
to experimental results as: mass of the Z-boson, $m_Z = 91.2 \gev$, 
top quark mass, $m_t = 173.2 \gev$, bottom quark mass, $m_b = 4.2
\gev$, the electromagnetic coupling, $\alpha_{em}^{-1} = 127.9$, and
the strong interaction coupling, $\alpha_s = 0.119$. The 
parameters are  
\bea \label{30parameters}
\underline{\theta} \equiv \{ \, 
M_{1,2,3}, \,\quad \mu, \,\quad m_A, \,\quad \tan \beta, \,\quad
Im(M_{1,2}, \,\quad \mu), \,\quad \tilde{a}_{1,2,\ldots,8}, 
Im(\tilde{a}_{4,5,8}), \,\quad x_{1,2}, \,\quad y_{1,3,4,5,6,7},
\,\quad Im(y_{4,5}) \, \}.
\eea
The MSSM-30 fits to the Higgs boson mass, the electroweak physics,
B-physics, lepton dipole moments and the cold dark matter relic
density observables disfavour large  $\tan \beta \gtrsim 30$. 
The corresponding posterior distribution on $(m_A, \tan \beta)$ plane
is show in \Figref{fig:mAtb}(a). The $(m_A, \tan \beta)$ plane is
chosen because we aim at showing the impact of the prior-independent
result \eq{cutfromfit} on the MSSM-30 posterior
sample. \Figref{fig:mAtb}(b) shows 
what remains after imposing the prior-independent naturalness
condition \eq{cutfromfit} by ruling out the unnatural points. From the
surviving posterior, it is deduced that $m_A \lesssim 3 \tev$ and $\tan \beta
\lesssim 25$ at 95\% Bayesian credibility. \footnote{Applying the naturalness
  line \eq{naturalness}, finding the allowed $m_A$ or $\tan \beta$
  95\% Bayesian credibility interval will require a separate fit of
  the MSSM-30 to data plus the naturalness line constraint beyond the
  scope of this article.} 

\paragraph{Collider limits on natural MSSM-30 points:}
To what extent does 8 TeV LHC probe the natural MSSM-30 posterior
point based on \Figref{cutfromfit}? The natural points can be checked
against some LHC limits. ATLAS
and CMS  95\% confidence level limits can be used to constrain
models that predict the processes they searched for. Limits on fiducial cross sections usually call for writing
Rivet~\cite{Buckley:2010ar} analyses to pass over
Herwig++~\cite{Bahr:2008pv} Monte Carlo generated supersymmetry
events. We did not intend to use the full set of such LHC
results. Rather, a selected few which are relevant for probing the
prior-independent naturalness condition \eq{cutfromfit} were considered. In \cite{Aad:2014ioa} a search for scalar 
particles decaying via narrow resonances into two photons is
performed. The limits 
applied on the MSSM-30 pseudoscalar Higgs production cross section
times branching ratio into two photons did not significantly
constrain the posterior sample. The ATLAS \cite{Aad:2014vgg} and CMS
\cite{Khachatryan:2014wca} limits from search for MSSM Higgs bosons
(here the pseudoscalar Higgs) 
decaying into tau-lepton pairs were also considered. These are put next to the production
cross section times branching fraction of the pseudoscalar decay into
tau-leptons for the MSSM-30 posterior computed using
FeynHiggs\cite{Heinemeyer:1998yj}. \Figref{fig:mAtb}(d) shows the
similar case for the ATLAS search for a CP-odd Higgs boson
decaying to the Z-boson and the SM Higgs boson which in turn decays to
tau-leptons \cite{Aad:2015wra}. All these searches hardly
constrain the natural MSSM-30 posterior mostly due to the low production
cross-section and decay rates of the pseudoscalar Higgs at the
LHC. Perhaps, searches with topologies involving the pseudoscalar MSSM
Higgs boson decaying via charginos and neutralinos could probe better
the naturalness allowed MSSM-30 region. 

\paragraph{Naturalness line:} Here we give a closer look at the
numerical and prior-independent result \eq{cutfromfit}. A zeroth-order
explanation for the bound on $m_A$ as function of $\tan \beta$ can be
explained using the electroweak fine-tuning measure 
$\Delta_{EW}$ \cite{Baer:2012up, Bae:2014fsa}. 
Consider the electroweak symmetry
breaking condition, assuming $\tan \beta >> 1$ but without lost of
generality  
\be \label{ewsb} \frac{1}{2} m_Z^2 \approx \frac{m_{H_d}^2}{\tan^2
  \beta} - m_{H_u}^2 -\mu^2. \ee  
{ Requiring there be no fine-tuning} will need all three terms on the
right hand side to be comparable amongst themselves and of order
$m_Z^2/2$. As such $\frac{m^2_{H_d}}{\tan^2 \beta} \sim m_Z^2/2$ and $-m^2_{H_u} - 
\mu^2 \sim 0$ implies that $m^2_{H_u} \sim -\mu^2$. In addition
$(\frac{m^2_{H_d}}{\tan^2 \beta})/(-m^2_{H_u}) \sim 1$ implies
$m^2_{H_d} >> - m^2_{H_u}$. Now applying $m^2_{H_u} \sim -\mu^2$ and
$m^2_{H_d} >> - m^2_{H_u}$ to the the tree-level relation $m_{A}^2 =
2|\mu|^2 + m_{H_d}^2 + m_{H_u}^2$ gives $m_{A}^2 \sim m_{H_d}^2.$ 
 Therefore requiring fine-tuning
$\Delta_{EW}$ no worst that $\Delta_{max}$ 
\be \label{DEW} \Delta_{EW} \equiv \frac{2 \, m_{H_d}^2}{m_Z^2 \,
  \tan^2 \beta} \leq \Delta_{max} \quad \implies \quad 
m_A < m_Z \, \tan \beta \, (\Delta_{max}/2)^{1/2}.
\ee

Loop corrections to the tree-level relation for $m_A^2$ is not going
spoil the bound \eq{DEW} or \eq{cutfromfit}. This is the case
for the loop-corrected \cite{Carena:2002es} $m_A$ used for the MSSM-25
fits as can be seen in \Figref{fig:mAtb_0}. There is also no
conflict with other fine-tuning measures
\cite{Harnik:2003rs,Kitano:2005wc,Ellis:1986yg,Barbieri:1987fn} since
all the measures agree with one another whenever appropriately applied
\cite{Baer:2014ica,Baer:2013gva}.  
As such the core message of this letter goes as follows. We seek
for robust predictions for assessing low-energy supersymmetry as the
model responsible for the Higgs boson mass stability. But this is not
possible as long as the subjectiveness inherent in the fine-tuning
measure \eq{DEW} 
remains. Constructing the bound in \eq{DEW} is based on the
{\it no fine-tuning} and {\it comparability} requirements for the terms in
\eq{ewsb}. Now imposing $\Delta_{EW} \leq \Delta_{max}$ is a
contradiction since {\it this allows fine-tuning} even if not worse
than $1/\Delta_{max}$. \Figref{fig:mAtb_0} give further insight to
this. The fits done with $\Delta_{max} = 20$ shows the corresponding
no-go regions similar to the case with $\Delta_{max} = 4$ which do not
agree with \eq{DEW}. Our take is that a model point is either
fine-tuned, meaning $\Delta_{EW} > 1$ or not fine-tuned when
$\Delta_{EW} = 1.$ This way {\it the subjectiveness in selecting a cut on
fine-tuning} is completely removed. The out come of this is a robust
``naturalness line'' 
\be
\label{naturalness} m_A \sim \frac{1}{\sqrt{2}} \, m_Z \, \tan \beta. 
\ee 
In fact imposing \eq{naturalness} reduces the $(m_A, \tan
\beta)$ plane into a line, meaning one less parameter in the Higgs
sector. Note that this result is not equivalent with the 
purely tree-level no fine-tuning measure $\mu \sim
\frac{1}{\sqrt{2}} \, m_Z$. The ansatz is that the naturalness line
holds at all loop levels such that radiative corrections to the masses
do not spoil the relation. The naturalness line \eq{naturalness} can
be used for mapping natural regions of any MSSM frame. 

\paragraph{Conclusions and outlook:}
We have addressed a question about finding an objective determinant
for the existence of natural supersymmetry. Our Bayesian approach
is based on finding prior-independent features within broader and
minimally biased frames as the guiding
principle~\cite{susy2013}. The results of this article and an outlook
are summarised as follows. 
\begin{itemize}
\item The 20-parameter MSSM fits to subjective
naturalness, using the electroweak fine-tuning measure, indicate the
existence of a prior-independent upper bound on the pseudoscalar Higgs
boson mass $m_A$ as a function of $\tan \beta.$ Imposing the bound on
the posterior sample of a 30-parameter MSSM fit to data shows that
$m_A \lesssim 3 \tev$ and $\tan\beta \lesssim 25$ at 95\% Bayesian
credibility region. The natural MSSM-30 points are not yet ruled out
by the 8 TeV LHC  limits we considered. Constraints from search
topologies that include decays into charginos and neutralinos could
lead to better probe.

\item We seek for robust predictions for assessing low-energy
supersymmetry as the model responsible for the Higgs boson mass
stability. We proposed that this is possible only if the
subjectiveness inherent in the electroweak fine-tuning measure is 
removed. Imposing 
$\Delta_{EW} \leq \Delta_{max}$ is a contradiction since {\it this
  allows fine-tuning} even if not worse than $1/\Delta_{max}$. A
robust method should require their either be fine-tuning, meaning
$\Delta_{EW} > 1$ or no fine-tuning, i.e. $\Delta_{EW} = 1.$ This way {\it
  the subjectiveness in selecting a cut on fine-tuning} is completely
removed and no fine-tuning means $m_A \sim \frac{1}{\sqrt{2}} \, m_Z
\, \tan \beta.$  We call this relation the ``naturalness line''.

\item ``Why is supersymmetry not yet discovered?'' Up to the public LHC
results as of the time of writing this article, we claim that an
answer is that  {\it the regions where it is expected are not yet
  probed.} ``Where to look for natural supersymmetry?'' The proposed
regions where it should be expected were derived   
via Bayesian method with minimal model framework construction or
theoretical prejudice. Natural supersymmetry should be looked for
along the ``naturalness line'' $m_A \sim \frac{1}{\sqrt{2}} \, m_Z \,
\tan \beta$ together with a 1-2 TeV lightest top-squarks. The 8 TeV
LHC limits on gluino and 1st-2nd generation sparticles are not in
conflict with these predictions.  

\item An interesting line for further studies will be to assess 
the impact of the full set of LHC fiducial cross section limits on the
``naturalness line'' in general and within particular phenomenological
frames such as the 30-parameters MSSM \cite{AbdusSalam:2014uea} of
the 2-parameters hMSSM \cite{Djouadi:2013uqa}.   
\end{itemize}

\paragraph{Acknowledgements:}
S.S. AbdusSalam is supported by funding from the European Research Council under 
the European Union's Seventh Framework Programme (FP/2007-2013)/ERC
Grant Agreement no. 279972 NPFlavour and would like to acknowledge the
hospitality from the IPM School of Particles and Accelerators, Tehran,
at some stage of the research presented. L. Velasco-Sevilla
acknowledges the support and hospitality from the ICTP, where part of
this work was carried out. 

\paragraph{Appendix: The expressions for the $\Sigma_u^u(\tilde
  t_{1,2}, \tilde b_{1,2})$ contributions to $\Delta_{EW}$} 
For self-sufficiency we explicitly show the expressions
for $\Sigma_u^u(\tilde t_{1,2}, \tilde b_{1,2})$ according to
\cite{Arnowitt:1992qp, Gladyshev:1996fx} 
\be
\Sigma_u^u(\tilde t_{1,2}) = \frac{3}{16\pi^2} \, F(m_{\tilde
  t_{1,2}}^2) \times \left[ f_t^2 - g_Z^2 \mp \frac{f_t^2
    A_t^2-8g_Z^2(\frac{1}{4}-\frac{2}{3}x_W)\Delta_t}{m_{\tilde t_2}^2
    - m_{\tilde t_1}^2} \right] \textrm{ and }
\ee
\be
\Sigma_d^d (\tilde t_{1,2}) = \frac{3}{16\pi^2} \, F(m_{\tilde
  t_{1,2}}^2) \left[ g_Z^2 \mp \frac{f_t^2\mu^2 + 8 g_Z^2 (\frac{1}{4}
    - \frac{2}{3} x_W) \Delta_t }{m_{\tilde t_2}^2 - m_{\tilde
      t_1}^2}\right]
\ee
where $\Delta_t=(m_{\tilde t_L}^2-m_{\tilde t_R}^2)/2 + M_Z^2 \, \cos
2\beta (\frac{1}{4} - \frac{2}{3} x_W)$, $g_Z^2=(g^2+g^{\prime 2})/8$,
$x_W \equiv \sin^2\theta_W$ and $F(m^2)=m^2\left(\log
(m^2/Q^2)-1\right)$, with $Q^2=m_{\tilde t_1}m_{\tilde
  t_2}$. $m_{\tilde t_{1,2}}$ are computed at tree-level. 
For the bottom-squarks,
\be
\Sigma_u^u (\tilde b_{1,2}) = \frac{3}{16\pi^2} \, F(m_{\tilde
  b_{1,2}}^2) \left[ g_Z^2 \mp \frac{f_b^2 \mu^2 - 8g_Z^2 (\frac{1}{4}
    - \frac{1}{3} x_W) \Delta_b}{m_{\tilde b_2}^2 - m_{\tilde
      b_1}^2}\right]  \textrm{ and } 
\ee
\be
\Sigma_d^d (\tilde b_{1,2}) = \frac{3}{16\pi^2} \, F(m_{\tilde
  b_{1,2}}^2) \left[ f_b^2 - g_Z^2 \mp \frac{f_b^2 A_b^2-8 g_Z^2
    (\frac{1}{4} - \frac{1}{3} x_W) \Delta_b}{m_{\tilde b_2}^2 -
    m_{\tilde b_1}^2}\right] 
\ee
where $\Delta_b=(m_{\tilde b_L}^2-m_{\tilde b_R}^2)/2 + M_Z^2 \, \cos
2\beta (\frac{1}{4} - \frac{1}{3} x_W)$. $m_{\tilde b_{1,2}}$ are
computed at tree-level.

\end{document}